\documentclass[aps,twocolumn,prl,tightenlines,floatfix,superscriptaddress]{revtex4-1}

\usepackage{graphicx}
\usepackage{pdfpages}
\usepackage{relsize}
\usepackage[breaklinks=true]{hyperref}
\usepackage[english]{babel}
\usepackage{amsmath}
\usepackage{amsfonts}
\usepackage{amssymb}
\usepackage{times}

\makeatletter
\AtBeginDocument{\let\LS@rot\@undefined}
\makeatother

\newcommand{\be}{\begin{equation}}
\newcommand{\ee}{\end{equation}}

\begin{document}

\title{Ultra high temperature superfluidity in ultracold atomic Fermi gases
  with mixed dimensionality}

\author{Leifeng Zhang }
\affiliation{Department of Physics and Zhejiang Institute of Modern
  Physics, Zhejiang University, Hangzhou, Zhejiang 310027, China}
\affiliation{Synergetic Innovation Center of Quantum Information and
  Quantum Physics, Hefei, Anhui 230026, China} 

\author{Jibiao Wang }
\affiliation{Laboratory of Quantum Engineering and Quantum Metrology, School of Physics and Astronomy, Sun Yat-Sen University (Zhuhai Campus), Zhuhai, Guangdong 519082, China} 

\author{Yi Yu }
\affiliation{Center for Measurements and Analyses, Zhejiang University of Technology, Hangzhou, Zhejiang 310014, China}
\affiliation{Department of Physics and Zhejiang Institute of Modern
  Physics, Zhejiang University, Hangzhou, Zhejiang 310027, China}

\author{Qijin Chen}
\email[Corresponding author: ]{qchen@uchicago.edu}
\affiliation{Department of Physics and Zhejiang Institute of Modern
  Physics, Zhejiang University, Hangzhou, Zhejiang 310027, China}
\affiliation{Synergetic Innovation Center of Quantum Information and
  Quantum Physics, Hefei, Anhui 230026, China} 

\date{\today}

\begin{abstract}
  Achieving a higher superfluid transition $T_c$ has been a goal for
  the fields of superconductivity and atomic Fermi gases. Here we
  propose that, by using mixed dimensionality, one may achieve ultra
  high temperature superfluids in two component atomic Fermi gases,
  where one component feels a regular three-dimensional (3D) continuum
  space, while the other is subject to a 1D optic lattice
  potential. Via tuning the lattice spacing and trap depth, 
  one can effectively raise the Fermi level dramatically upon pairing
  so that superfluidity may occur at an ultra high temperature (in
  units of Fermi energy) even beyond the quantum degeneracy regime,
  well surpassing that in an ordinary 3D Fermi gas and all other known
  superfluids and superconductors.
\end{abstract}


\maketitle

It has been an important goal to achieve higher or even room
temperature superconductivity \cite{Chu2013conf}, since the discovery
of high $T_c$ superconductors in 1986 \cite{Bednorz}, with a typical
maximum transition temperature $T_c$ of around 95~K for optimally
doped YBa$_2$Cu$_3$O$_{7-\delta}$ and
Bi$_{2}$Sr$_{2}$CaCu$_{2}$O$_{8+\delta}$ \cite{Timusk}. However,
except for the Hg-based HgBa$_2$Ca$_2$Cu$_3$O$_{8+\delta}$ (for which
$T_c$ can be up to 164~K) under high pressure \cite{Chu1994}, there
has been essentially not much progress in raising $T_c$. The typical
$T_c/T_F$ is only around 0.05 or less.
There have been a few other families of superconductors, besides
conventional metal superconductors and the cuprates. These include
iron-based superconductors \cite{Hosono}, heavy fermion
superconductors \cite{Steglich1979} and organic superconductors
\cite{McKenzie}. Despite the similarity in phase diagrams among these
different families, the maximum attainable $T_c/T_F$ has not been able
to exceed that of the cuprates.

Other notable superconductors include the recently discovered H$_2$S
with a record high $T_c=203$~K, which however requires an enormous
high pressure of 90 GPa \cite{H2S}, and the monolayer FeSe/SrTiO$_3$
superconductors with a $T_c$ (or gap opening temperature) up to 100K
\cite{Xue2012,Monolayer100K}.  The suggested conventional
electron-phonon based pairing mechanism for both systems
\cite{H2S,LeeDH} implies that their $T_c/T_F$ is very low.
The very recently discovered superconductivity in twisted angle
bilayer graphenes \cite{YCao1} has $T_c = 1.7$K with a near-flat
band-width of 10 meV, leading to $T_c/T_F \sim 0.04$ (here we take
$T_F = 5$ meV/$\hbar$), comparable to the cuprates.

There have been indications \cite{Uemura} for connections between high
$T_c$ superconductivity and BCS--Bose-Einstein condensation (BEC)
crossover, the latter of which has a the BEC asymptote of
$T_c = 0.218 T_F$ in three-dimensional (3D) continuum, where $T_F$
denotes the Fermi temperature. It has now been clear that with a
$d$-wave pairing symmetry, high $T_c$ cuprates cannot reach the BEC
regime \cite{Chen1}; instead, they fall in between the BCS and BEC
regimes, with a strong signature of a pseudogap in the single-particle
excitation spectrum, which has been referred to as a pseudogap or
crossover regime.

With the advent of superfluidity in ultracold atomic Fermi gases, the
hope for achieving a higher $T_c/T_F$ in these systems arose. Indeed,
BCS-BEC crossover in 3D Fermi gases has been realized experimentally
since 2004 \cite{Review}.  While a substantially higher $T_c/T_F^0$ is
possible in a trap in the BEC regime, (with a BEC asymptote of 0.518,
where $T_c^0$ is the noninteracting global Fermi temperature), the
maximum $T_c$ in a homogeneous system occurs in the vicinity of
unitarity, inside the pseudogap or crossover regime. Its value varies
around $T_c/T_F \sim 0.2$ in different theoretical calculations
\cite{NSR,Zwerger} and quantum Monte Carlo simulations
\cite{TroyerPRL2008,Wingate,BulgacMC,Akkineni,Floerchinger} as well as
experimental measurements \cite{ThermoScience,Zwierlein2012}. Indeed,
the local $T_c(\mathbf{r})/T_F(\mathbf{r})$ in the trap never exceeds
that of a homogeneous system. The increased BEC asymptote for $T_c$ is
a consequence of an increased local density (and hence local Fermi
energy) at the trap center.

Fermions on a lattice have also been intensively investigated
theoretically. However, the maximum $T_c/T_F$ cannot surpass their
continuum counterpart, since the lattice periodicity usually has a
negative impact on the fermion mobility and thus serves to suppress
$T_c$ \cite{Chen1}.

In this Letter, we propose that using an artificially engineered mixed
dimensional setting, one may achieve ultra high superfluid transition
temperature $T_c$ in units of $T_F$. We find that, owing to the
special features of the mixed dimensions, one can maintain a high
$T_c$,  unsuppressed by a tiny lattice hopping integral $t$.
We show that the maximum attainable $T_c/T_F$ may reach unity or even
beyond the quantum degeneracy regime, well surpassing the maximum
values in a pure dimensional system or any known superfluids. This may
shed light in the ultimate search for room temperature
superconductivity.

Mixed dimensionality is realizable experimentally.  Recently,
Lamporesi \textit{et al.}  \cite{Lamporesi10PRL} has successfully
obtained a mixed-dimensional system with a Bose-Bose mixture of
$^{41}$K--$^{87}$Rb; only $^{41}$K atoms feel the lattice potential,
leaving $^{87}$Rb atoms moving freely in the 3D continuum. The species
selective technique for the optical potential is also applicable to
fermionic atoms.  Therefore, one may realize mixed dimensions for
atomic Fermi gases as well.

On the theory side, mixed dimensionality has been of interest since
the pioneering work of Iskin and coworkers \cite{Iskin10PRA}, who
investigated the phase diagrams of equal population fermion mixtures
in the framework of BCS--BEC crossover at zero temperature $T$ in
mixed dimensions, using a strict mean-field approach. A preliminary
study of finite temperature cases was reported
\cite{XYang11EPJB}. Recently, a more systematic investigation of the
pairing and superfluid phenomena at finite temperatures in mixed
dimensions has been reported for an equal mass and equal population
case \cite{MixedDim_Srep}. The result seems to suggest that $T_c$ is
substantially higher for the cases of a larger lattice spacing
$d$. However, one may also notice that these large $d$ situations are
not readily accessible in simple experiments.  The potential to
achieve a higher $T_c$ within a physically accessible range of tuning
parameters using mixed dimensionality needs more careful
investigations.

Here we explore the effects of mixed dimensionality on the enhancement
of Fermi level and show how this may lead to ultra high superfluid
transition temperatures $T_c/T_F$ in two-component atomic Fermi gases.
Due to the high complexity caused by multiple tunable parameters, here
we restrict ourselves to the population \emph{balanced} case with
equal masses, and avoid other complications that may arise from
possible Fulde-Ferrell-Larkin-Ovchinnikov (FFLO) states \cite{FF,LO}
and phase separation, which can exist only at low $T$ and are thus
irrelevant at relatively high temperatures \cite{FFLO_Instability}.

We shall consider the same dimensional setting as in the experiment of
Ref.~\cite{Lamporesi10PRL}, and use the same formalism based on a
pairing fluctuation theory \cite{Chen2,Review} as presented in
Ref.~\cite{MixedDim_Srep} for mixed dimensions. We refer to the
lattice and 3D continuum components as spin up and spin down,
respectively, and define the Fermi energy naturally as
$E_F = \hbar^2 k_F^2/(2m)$, with $k_F^{}=(6\pi^2 n_\downarrow)^{1/3}$
being the Fermi momentum of the 3D component (we have set
$\hbar=1$).

To keep the paper self-contained, we recapitulate the formalism.  The
band dispersions for the lattice and the 3D components are given by
$\xi_{\mathbf{k}\uparrow}=\mathbf{k}_\parallel^2/2m+2t[1-\cos(k_zd)]-\mu_\uparrow$
and $\xi_{\mathbf{k}\downarrow}=\mathbf{k}^2/2m-\mu_\downarrow$,
respectively. Here $\mathbf{k}_\parallel\equiv(k_x,k_y)$, where
$\mu_\sigma$ (with $\sigma=\uparrow,\downarrow$) are the fermionic
chemical potentials, 
and $t$ is the hopping integral between nearest neighbor sites in the
lattice dimension.  The one-band assumption is appropriate when the
lattice band gap is experimentally tuned to be large compared with
Fermi energy $E_F$. Pairing takes place via an $s$-wave short range
attractive interaction.

Both superfluid condensate, if present, and noncondensed pairs
contribute to the fermion self-energy, and thus to the single particle
excitation gap $\Delta$, via $\Delta^2=\Delta_{sc}^2+\Delta_{pg}^2$,
where $\Delta_{sc}$ and $\Delta_{pg}$ are the superfluid order
parameter and the pseudogap, respectively. Using the same four-vector
notations as in Refs.~\cite{Review,MixedDim_Srep}, the full Green's
functions are given by
\begin{eqnarray}
  G_\sigma(K)&=&\frac{u_\mathbf{k}^2}{i\omega_n-E_{\mathbf{k}\sigma}}
             +\frac{v_\mathbf{k}^2}{i\omega_n+E_{\mathbf{k}\bar{\sigma}}},\quad
             |k_z|<\frac{\pi}{d}\nonumber\\
G_\downarrow(K)&=&\frac{1}{i\omega_n-\xi_{\mathbf{k}\downarrow}}, \quad
|k_z|>\frac{\pi}{d}\,
  \label{eq:GF}
\end{eqnarray}
where
$u_\mathbf{k}^2=(1+\xi_\mathbf{k}/E_\mathbf{k})/2,
v_\mathbf{k}^2=(1-\xi_\mathbf{k}/E_\mathbf{k})/2$,
$E_\mathbf{k}=\sqrt{\xi_\mathbf{k}^2+\Delta^2}$, and
$E_{\mathbf{k}\sigma}=E_\mathbf{k}+\zeta_{\mathbf{k}\sigma}$,
$\xi_\mathbf{k}=(\xi_{\mathbf{k}\uparrow}+\xi_{\mathbf{k}\downarrow})/2$,
$\zeta_{\mathbf{k}\sigma}=(\xi_{\mathbf{k}\sigma}-\xi_{\mathbf{k}\bar{\sigma}})/2$. Note
that we have neglected the finite-momentum pairing effects on
$G_\downarrow(K)$ outside the first Brillouin zone (BZ).

The equations for the total atomic number density
$n=n_\uparrow+n_\downarrow$ and the number difference
$\delta n=n_\uparrow-n_\downarrow = 0$ are given by
\begin{eqnarray}
  n&=&2\sum_\mathbf{k}\left[
      v_\mathbf{k}^2
      +\bar{f}(E_\mathbf{k})\frac{\xi_\mathbf{k}}{E_\mathbf{k}}
    \right]
    +\!\!\!\!\sum_{|k_z|>{\pi}/{d}}\!\!\!\! f(\xi_{\mathbf{k}\downarrow})\,,
  \label{eq:ntot}
\\
0&=&\sum_\mathbf{k}\left[
    f(E_{\mathbf{k}\uparrow})-f(E_{\mathbf{k}\downarrow})
  \right]
  -\!\!\!\!\sum_{|k_z|>{\pi}/{d}}\!\!\!\! f(\xi_{\mathbf{k}\downarrow})\,,
  \label{eq:dn}
\end{eqnarray}
where $f(x)$ is the Fermi distribution function, and the average
$\bar{f}(x)\equiv{\sum_\sigma f(x+\zeta_{\mathbf{k}\sigma})/2}$.

Similar to the pure 3D case, an $s$-wave scattering length $a$ in
mixed dimensions is defined via the Lippmann-Schwinger relation
$g^{-1}=m/4\pi a-\sum_\mathbf{k}1/2\epsilon_\mathbf{k}$, where
$\epsilon_\mathbf{k}=(\epsilon_{\mathbf{k}\uparrow} +
\epsilon_{\mathbf{k}\downarrow})/2$,
with $\epsilon_{\mathbf{k}\sigma}=\xi_{\mathbf{k}\sigma}+\mu_\sigma$,
and $g<0$ is the pairing strength.  In the superfluid state, the
Thouless criterion leads to the gap equation
\begin{equation}
  \frac{m}{4\pi a}=\dfrac{m_{eff}}{4\pi a_{eff}} =\sum_\mathbf{k}\left[
    \frac{1}{2\epsilon_\mathbf{k}}-\frac{1-2\bar{f}(E_\mathbf{k})}{2E_\mathbf{k}}
  \right]\,,
\label{eq:gap}
\end{equation}
where the 3D equivalent effective mass, $m_{eff}$, which better
reflects the lattice contribution, can be deduced from the trace of
the inverse mass tensor \cite{noteonm_eff},
$\dfrac{1}{m_{eff}} = \dfrac{5}{6m}+\dfrac{1}{3}td^2 $.
This then defines an effective scattering length $a_{eff}$
such that
$\dfrac{1}{k_F^{}a_{eff}} = \dfrac{1}{k_F^{}a} \left(\dfrac{5}{6} +
  \dfrac{m}{3}td^2\right)$.
The quantity $a_{eff}$ reflects the actual scattering length that can
be measured experimentally \cite{Lamporesi10PRL}. A plot of
$a/a_{eff}$ as a function of $k_F^{}d$ for $t/E_F=0.1$ is shown in
Supplementary Fig.~S1.

The pair dispersion can be deduced via Taylor expanding the inverse
$T$ matrix as
$t_{pg}^{-1}(Q)\approx{Z_1 (i\Omega_l)^2 +
  Z(i\Omega_l-{\Omega}_\mathbf{q})}$,
where ${\Omega}_\mathbf{q}=q_\parallel^2/2M_\parallel^*+q_z^2/2M_z^*$
in the superfluid phase \cite{Review}, with anisotropic effective pair
masses $M_\parallel^*$ and $M_z^*$ in the in-plane and out-of-plane
directions, respectively. The coefficients $Z$, $Z_1$, $1/M_\parallel$
and $1/M_z$ can be computed during the expansion.

The pseudogap $\Delta_{pg}$ is related to the density of pairs, via
\begin{equation}
  \Delta_{pg}^{2}=\sum_{\mathbf{q}_\parallel^{}}\sum_{|q_z|<\pi/d}\frac{b(\tilde{\Omega}_{\textbf{q}})}
  {Z\sqrt{1+4\dfrac{Z_{1}}{Z}  {\Omega}_\mathbf{q}}}\,,
  \label{eq:pg}
\end{equation}
where $b(x)$ is the Bose distribution function and
$\tilde{\Omega}_{\textbf{q}}=Z\{\sqrt{1+4Z_{1}
  {\Omega}_{\textbf{q}}/Z}-1\}/2Z_{1}$
is the pair dispersion.

The closed set of equations~(\ref{eq:ntot})--(\ref{eq:pg}) will be
used to solve for $T_c$ (and the pseudogap $\Delta_{pg}$ and chemical
potentials at $T_c$), by setting the order parameter
$\Delta_{sc}=0$.

\begin{figure}
  \centerline{\includegraphics[clip,width=3.2in] {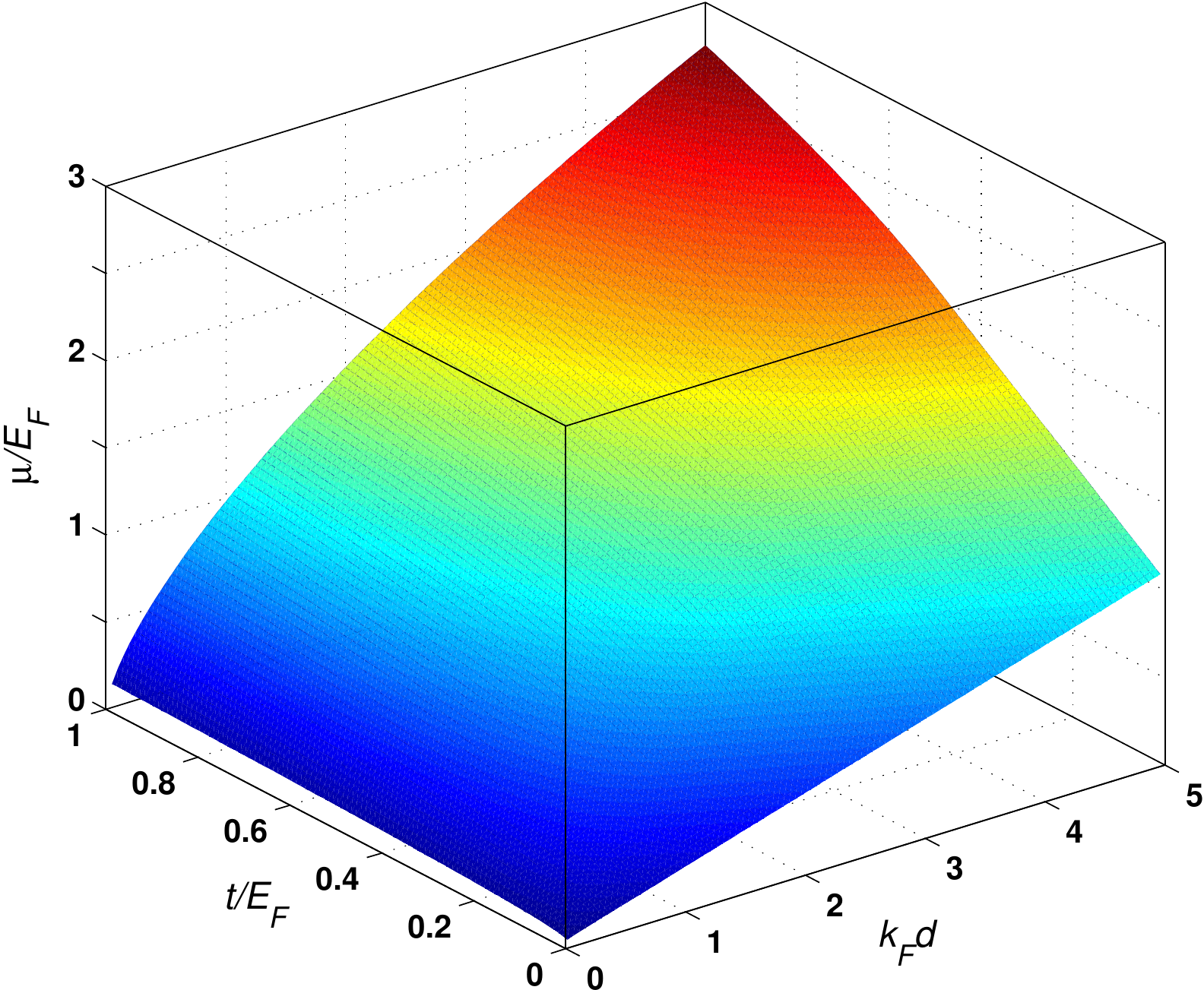}}
  \caption{ Evolution of the chemical potential $\mu_\uparrow$ of the
    lattice component, as a function of $t$ and $d$. $\mu_\uparrow$
    stays low for small $d$ and becomes elevated for large $d$. }
\label{fig:mu}
\end{figure}

The solution for $T_c$ in the deep BEC regime can be simplified
dramatically, where everything is small compared with $|\mu|$. It can
be shown that the $Z_1\Omega^2$ term is negligible, and
$M_\parallel^* = M_z^* = 2m$ so that
$\Omega_\mathbf{q} = \frac{q^2}{4m}$. The only equation that matters
for $T_c$ is the pseudogap equation (\ref{eq:pg}), which then reduces
to
\begin{equation}
\frac{n}{2} = \sum_\mathbf{q_\parallel}\sum_{|q_z|\le\pi/d} b(\Omega_\mathbf{q})\,.
\end{equation}
From this equation, we can see that the BEC asymptote for $T_c$ does
depends on $d$, and this dependence becomes stronger as $d$ becomes
larger. A larger $d$ means a more reduced phase space in
the $\hat{z}$-direction, and thus needs a higher $T_c$ to excite pairs into
higher $q_\parallel$ states, in order to satisfy the boson number
conservation.

To substantiate our idea of pushing up the Fermi level using mixed
dimensions, we present in Fig.~\ref{fig:mu} the evolution of chemical
potential $\mu_\uparrow$ of the lattice component in the
noninteracting limit as a function of $t$ and $d$. It is evident that
$\mu_\uparrow$ increases monotonically with both increasing $d$ and
$t$. While a large $t$ is unphysical, we shall focus mainly on the
effect of increasing $d$. This elevated Fermi level $\mu_\uparrow$ can
be understood from the Fermi ``disk''-like filling in momentum space,
as exemplified by Supplementary Fig.~S2 for $t=0.01E_F$ and
$k_F^{}d=8$.  For large $d$, the $k_z$ levels are limited, so that
particles are forced to occupy high $k_\parallel$ levels, leading to
an elevated Fermi level. Furthermore, in a simple lattice, the hopping
integral $t$ decreases with increasing $d$. Despite the fact that one
may tune the shape of the lattice potential to maintain a relatively
large $t$, the product $td^2$ is upper bounded by $1/2m$, which
corresponds to the zero lattice depth limit.  For this reason, we
shall keep the product $td^2$ small while changing $d$.

\begin{figure}
  \centerline{\includegraphics[clip,width=3.3in] {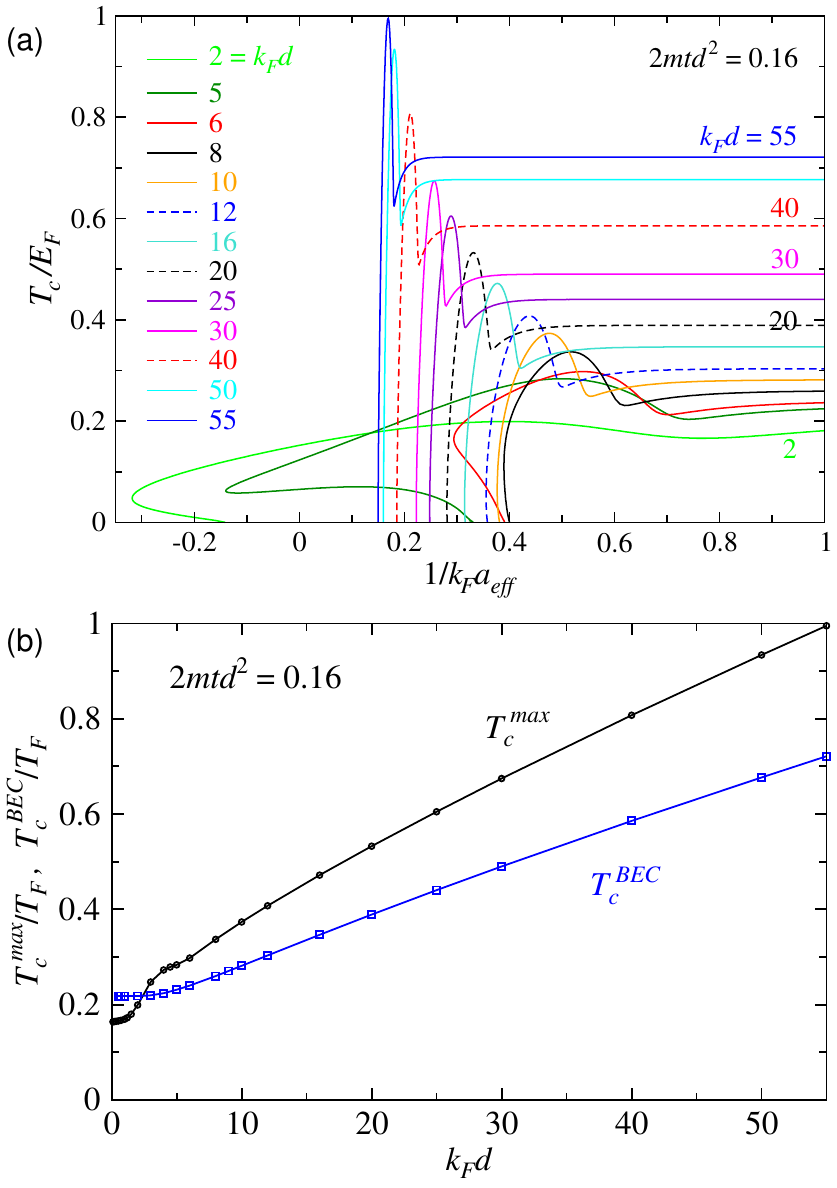}}
  \caption{(Color online) Effect of a large $d$ on the maximum
    $T_c$. (a) Behavior of $T_c$ as functions of
    $1/k_F^{}a_{eff}$ at fixed $2mtd^2 = 0.16$, but for different
    values of $k_F^{}d$ from 1 up to 55. (b) Plot of
    $T_c^{max}/T_F^{}$ and $T_c^{BEC}/T_F^{}$ as a function of $k_F^{}d$.}
\label{fig:Tc-d} 
\end{figure}

Our main result is presented in Fig.~\ref{fig:Tc-d}. Here we show in
Fig.~\ref{fig:Tc-d}(a) a series of $T_c$ curves as a function of
$1/k_F^{}a_{eff}$ at a realistic value of $2mtd^2 = 0.16$, but for
different values of $k_F^{}d$ from 1 up to 55.  Each curve has a
maximum $T_c$, $T_c^{max}$, and a BEC asymptote $T_c^{BEC}$ in the
large $1/k_F^{}a_{eff}$ limit. As $d$ increases, both $T_c^{max}$ and
$T_c^{BEC}$ increase progressively. For $k_F^{}d=55$, we have
$T_c^{max} = 0.995 \approx 1$. In Fig.~\ref{fig:Tc-d}(b), we plot
$T_c^{max}$ and $T_c^{BEC}$ as a function of $d$. Figure
\ref{fig:Tc-d} indicates that $T_c^{max}$ and $T_c^{BEC}$ increase
with $k_F^{}d$ almost linearly, without an upper bound. At
$k_F^{}d = 55$, the maximum $T_c$ is close to $T_F^{}$, and the BEC
asymptote $T_c^{BEC}$ has risen up to $0.72T_F^{}$. As a
self-consistency check, we note that $T_c^{BEC}/T_F$ approaches the
pure 3D value, 0.218, when $d$ decreases below $\pi$. Other quantities
including excitation gap, pairing strength and chemical potentials at
the maximum $T_c$ points are plotted in Supplementary Fig.~S3. Note
that for fixed $td^2$, the relation $a/a_{eff}\approx 0.86$ is the
same for all curves in Fig.~\ref{fig:Tc-d}(a).

One my notice that $T_c$ divided by the noninteracting $\mu_\uparrow$
from Fig.~1 will not exhibit such an increase with $d$, since the
noninteracting $\mu_\uparrow$ is roughly linear in $d$. However, we
argue that, unlike a pure lattice case, the mixed dimensional sitting
provides the noninteracting $E_F$ of the 3D component as a natural
energy scale, so that the noninteracting $\mu_\uparrow$ becomes a
variable that can be tuned via $t$ and $d$. Alternatively, one may
think of the increase of $T_c$ as compared with a given Fermi gas of
the same atom density in 3D continuum.

\begin{figure}
  \centerline{\includegraphics[clip,width=3.2in] {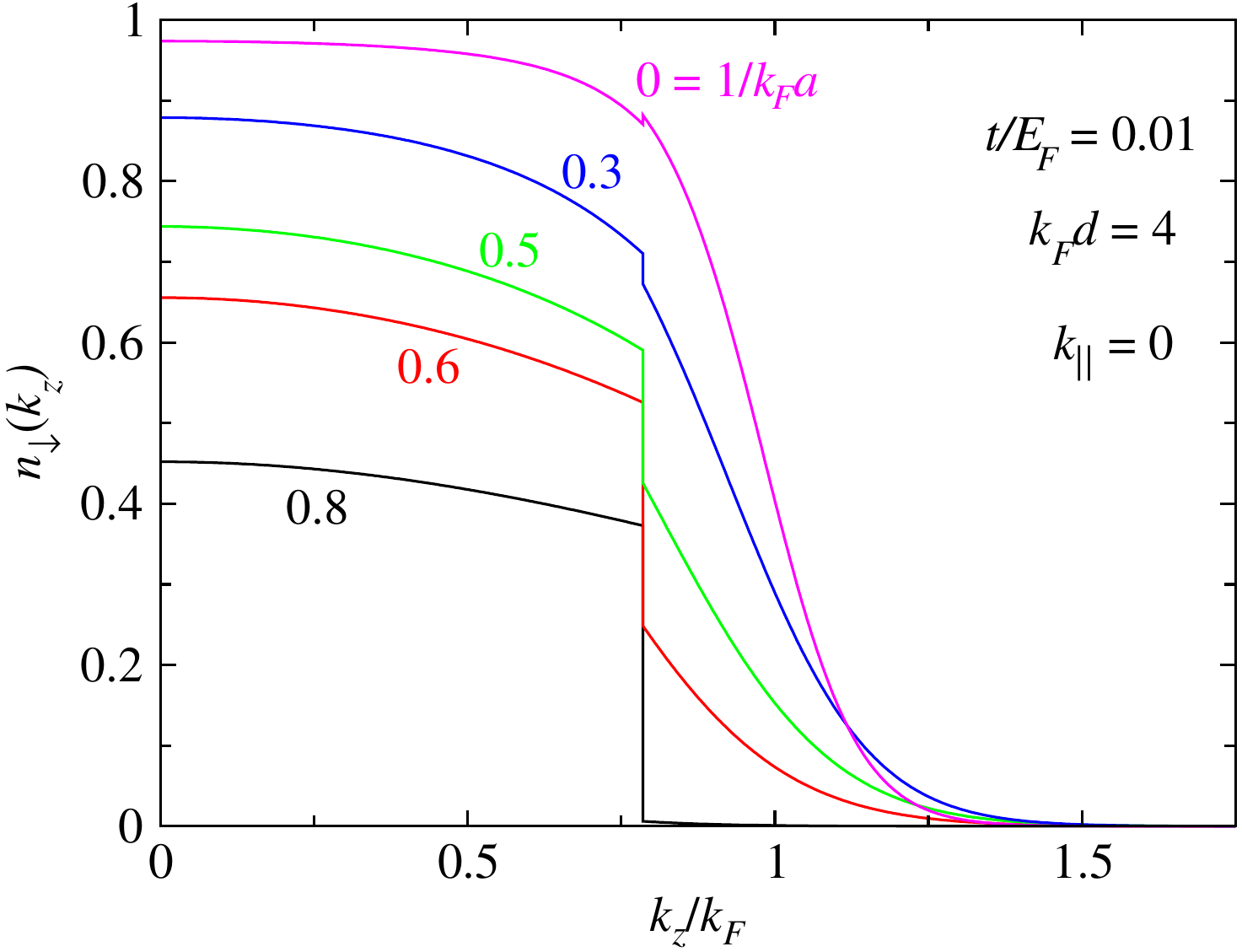}}
  \caption{Momentum distribution $n(\mathbf{k}_\parallel=0,k_z)$ of
    the 3D component along the $k_z$ axis for different pairing
    strength, characterized by $1/k_F^{}a$, with $t/E_F = 0.01$ and
    $k_F^{}d=4$. Here we fix the in-plane momentum
    $\mathbf{k}_\parallel = 0$. Upon entering the BEC regime, the
    occupation for $k_z > \pi/d$ decreases rapidly. }
\label{fig:momentumdistrib}
\end{figure}

To further ascertain the evolution of the Fermi level at large $d$, we
investigate the momentum distributions for different pairing
strengths. Shown in Fig.~\ref{fig:momentumdistrib} is the momentum
distribution $n_\downarrow(\mathbf{k}_\parallel=0,k_z)$ of the 3D
components for $k_F^{}d=4$ and $t/E_F = 0.01$ along the $k_z$ axis with
different pairing strengths in the unitary and near BEC regimes. As
$1/k_F^{}a$ increases from unitarity, the spectral weight outside the
first BZ decreases rapidly, and essentially vanishes for
$1/k_F^{}a=0.8$.

The corresponding in-plane momentum distribution
$n_\downarrow(\mathbf{k}_\parallel,k_z=0)$ of the 3D component in the
$k_z=0$ plane, as shown in Supplementary Fig.~S4, does not look
qualitatively much different from its pure 3D counterpart. The lack of
sharp features makes it hard to discern by naked eyes the changes in
$n_\downarrow(\mathbf{k}_\parallel,k_z=0)$ caused by a shrinking
distribution in the $k_z$ direction.

An inspection of $n_\downarrow(\mathbf{k}_\parallel=0,k_z)$, the
momentum distribution of the 3D component along the $\hat{k}_z$ axis,
at the maximum $T_c$ points for a series of $d$ values, as shown in
Supplementary Fig.~S5, reveals that the spectral weight outside the
first BZ is not necessarily zero in order to reach the maximum $T_c$;
there is still a considerable mismatch in momentum distributions
between the two pairing components at these maximum $T_c$ points.  The
corresponding in-plane momentum distribution is shown in Supplementary
Fig.~S6(a). 
The shift of the spectral weight toward higher $k_\parallel$ with
increasing $d$ can be made apparent through the higher order moments,
$ k_\parallel^n n_\downarrow(\mathbf{k}_\parallel,k_z=0)$.  As shown
in Supplementary Fig.~S6(b) for $n=2$, both the peak location and the
peak height increase with $d$.

\begin{figure}
  \centerline{\includegraphics[clip,width=3.2in] {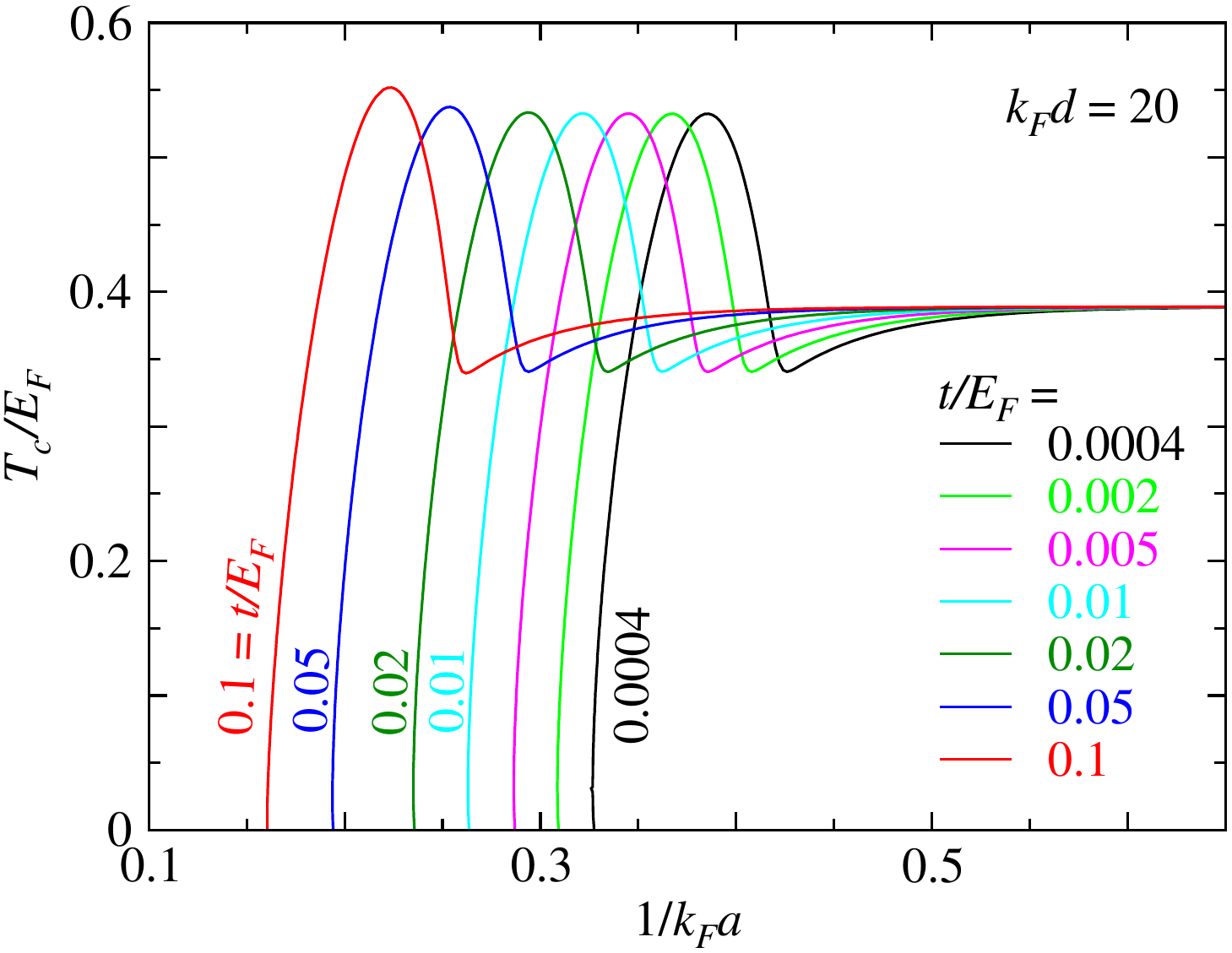}}
  \caption{Effects of $t$ on the behavior of $T_c$ for fixed $k_F^{}d=20$
    with different $t/E_F$ from 0.00004 up to 0.1.  }
\label{fig:Tc-t}
\end{figure}

We also investigate the effect of a varying hopping matrix element $t$
on $T_c$ with a fixed $d$. For small $d$, $t$ can vary over a
relatively large range. However, for large $d$, the experimentally
accessible range of $t$ is fairly small. Shown in Fig.~\ref{fig:Tc-t}
are the $T_c$ curves versus $1/k_F^{}a$ for $k_F^{}d = 20$ with
different $t/E_F^{}$ from 0.00004 up to 0.1. Note that except for the
lowest two $t$ values, the rest curves are not readily accessible in
experiment \cite{noteonlarge_td2}. Nonetheless, these $T_c$ curves
look like a parallel shift of one another, indicating that besides this
shift, $T_c$ barely changes over this range of small $t$. Further
increase of $t$ would lead to a more pronounced increase in $T_c$.
A replot of Fig.~\ref{fig:Tc-t} as a function of $1/k_Fa_{eff}$ is
given in Supplementary Fig.~S7, where the $T_c$ curves are
horizontally rescaled by different factors.

We note that it may not be easy to control very large $d$ values
experimentally. In addition, the pairing gap at the maximum $T_c$ for
the $d=50$ case is huge, as shown in Supplementary Fig.~S3(a). This
likely points to the need to include higher energy bands in the
lattice dimension. Nevertheless, we argue that as long as the band gap
is large, the contributions from the higher energy bands will only
cause a secondary, quantitative correction to $T_c$. It shall remain
valid that a large $d$ in the mixed dimensions will substantially
enhance $T_c$.

Experimentally a Fermi-Fermi mixture may be needed in order to achieve
mixed dimensions.  Nonetheless, the Fermi momentum does not depend on
the atomic mass $m$ or hopping integral $t$. Therefore, upon pairing,
one can still achieve a perfect Fermi surface match, as long as the
populations are balanced \cite{noteonmassimb}.
The mechanism for the enhancement of $T_c$ is still valid. A close
match between masses may occur for pairing between two isotopic
fermionic atoms, such as $^{161}$Dy and $^{163}$Dy.  Detailed
quantitative influences of a mass imbalance (and other factors such as
dipolar interactions) will be investigated in future studies.

It should be emphasized that \emph{our findings about the enhancement
  of $T_c$ via mixed dimensions are essentially independent of the
  details of our pairing fluctuation theory}. Alternative theories
such as the Nozeres--Schmit-Rink \cite{NSR} and FLEX approximations
\cite{Bickers} of the $T$-matrix theories should yield
\emph{qualitatively} similar results.

Finally we note that a key difference between the mixed dimensions and
pure lattice cases is that the effective pair mass $M_z$ in the
lattice direction is drastically different in the BEC asymptote. For
the former case, $M_z$ is shown to be equal to $2m$, since the total
kinetic energy of the two pairing atoms is dominated by the 3D
component in such a way that the pairs never become local around the
lattice sites. This is an unusual feature of the mixed
dimensionality. In contrast, in the pure lattice case, pairs move
mainly via virtual ionization \cite{NSR}. This leads to an effective
pair hopping integral $t_B\sim -t^2/g$ that decreases with increasing
pairing strength $|g|$, so that
$M_z \sim 1/t_B \sim |g|$ becomes heavy in the BEC regime.

We emphasize that \emph{this difference is the key} to understand why for the
tiny $t/E_F=5.3\times 10^{-5}$ in the case of $k_F^{}d=55$, the $T_c$
solution is not strongly suppressed.  This also implies that the BEC
asymptote $T_c^{BEC}$ is governed by $d$, whereas $t$ becomes totally
irrelevant.

In summary, we have studied the enhancement effect of a large $d$ (and
$t$) on the behavior of $T_c$ in mixed dimensions using a pairing
fluctuation theory. We propose that one may achieve ultra high
temperature superfluids using such a mixed dimensional setting with a
large $d$. A strong pairing interaction may bring all fermions of the
3D component to within the first Brillouin zone in the lattice
direction so that the Fermi level is pushed up. As a consequence, this
leads to a greatly enhanced $T_c$, all the way up to (or even beyond)
the quantum degeneracy temperature $T_F^{}$.  Unlike an pure optical
lattice case, the BEC asymptote for $T_c$ is pushed up dramatically as
well.  These predictions can be tested experimentally.
How to extend current results to condensed matter systems will be an
interesting subject for future studies, which should shed light on the
search for room temperature superconductors.

\begin{acknowledgments}

  We thank useful discussions with Brandon Anderson, Yuao Chen, Kuiyi
  Gao, A.J. Leggett, K. Levin, Gentaro Watanabe, Bo Yan and Xingcan
  Yao. This work is supported by NSF of China (Grants No.  11274267
  and No. 11774309), the National Basic Research Program of China
  (No. 2012CB927404), NSF of Zhejiang Province of China (Grant No.
  LZ13A040001).  Part of this work was completed while QC was visiting
  the University of Chicago.
\end{acknowledgments}


\begin{thebibliography}{37}%
\makeatletter
\providecommand \@ifxundefined [1]{%
 \@ifx{#1\undefined}
}%
\providecommand \@ifnum [1]{%
 \ifnum #1\expandafter \@firstoftwo
 \else \expandafter \@secondoftwo
 \fi
}%
\providecommand \@ifx [1]{%
 \ifx #1\expandafter \@firstoftwo
 \else \expandafter \@secondoftwo
 \fi
}%
\providecommand \natexlab [1]{#1}%
\providecommand \enquote  [1]{``#1''}%
\providecommand \bibnamefont  [1]{#1}%
\providecommand \bibfnamefont [1]{#1}%
\providecommand \citenamefont [1]{#1}%
\providecommand \href@noop [0]{\@secondoftwo}%
\providecommand \href [0]{\begingroup \@sanitize@url \@href}%
\providecommand \@href[1]{\@@startlink{#1}\@@href}%
\providecommand \@@href[1]{\endgroup#1\@@endlink}%
\providecommand \@sanitize@url [0]{\catcode `\\12\catcode `\$12\catcode
  `\&12\catcode `\#12\catcode `\^12\catcode `\_12\catcode `\%12\relax}%
\providecommand \@@startlink[1]{}%
\providecommand \@@endlink[0]{}%
\providecommand \url  [0]{\begingroup\@sanitize@url \@url }%
\providecommand \@url [1]{\endgroup\@href {#1}{\urlprefix }}%
\providecommand \urlprefix  [0]{URL }%
\providecommand \Eprint [0]{\href }%
\providecommand \doibase [0]{http://dx.doi.org/}%
\providecommand \selectlanguage [0]{\@gobble}%
\providecommand \bibinfo  [0]{\@secondoftwo}%
\providecommand \bibfield  [0]{\@secondoftwo}%
\providecommand \translation [1]{[#1]}%
\providecommand \BibitemOpen [0]{}%
\providecommand \bibitemStop [0]{}%
\providecommand \bibitemNoStop [0]{.\EOS\space}%
\providecommand \EOS [0]{\spacefactor3000\relax}%
\providecommand \BibitemShut  [1]{\csname bibitem#1\endcsname}%
\let\auto@bib@innerbib\@empty
\bibitem [{\citenamefont {Chu}\ \emph {et~al.}(2013)\citenamefont {Chu},
  \citenamefont {Lv}, \citenamefont {Deng}, \citenamefont {Lorenz},
  \citenamefont {Jawdat}, \citenamefont {Gooch}, \citenamefont {Shrestha},
  \citenamefont {Zhao}, \citenamefont {Zhu}, \citenamefont {Xue},\ and\
  \citenamefont {Wei}}]{Chu2013conf}%
  \BibitemOpen
  \bibfield  {author} {\bibinfo {author} {\bibfnamefont {C.~W.}\ \bibnamefont
  {Chu}}, \bibinfo {author} {\bibfnamefont {B.}~\bibnamefont {Lv}}, \bibinfo
  {author} {\bibfnamefont {L.~Z.}\ \bibnamefont {Deng}}, \bibinfo {author}
  {\bibfnamefont {B.}~\bibnamefont {Lorenz}}, \bibinfo {author} {\bibfnamefont
  {B.}~\bibnamefont {Jawdat}}, \bibinfo {author} {\bibfnamefont
  {M.}~\bibnamefont {Gooch}}, \bibinfo {author} {\bibfnamefont
  {K.}~\bibnamefont {Shrestha}}, \bibinfo {author} {\bibfnamefont
  {K.}~\bibnamefont {Zhao}}, \bibinfo {author} {\bibfnamefont {X.~Y.}\
  \bibnamefont {Zhu}}, \bibinfo {author} {\bibfnamefont {Y.~Y.}\ \bibnamefont
  {Xue}}, \ and\ \bibinfo {author} {\bibfnamefont {F.~Y.}\ \bibnamefont
  {Wei}},\ }\href {http://stacks.iop.org/1742-6596/449/i=1/a=012014} {\bibfield
   {journal} {\bibinfo  {journal} {J. Phys.: Conf. Series}\ }\textbf {\bibinfo
  {volume} {449}},\ \bibinfo {pages} {012014} (\bibinfo {year}
  {2013})}\BibitemShut {NoStop}%
\bibitem [{\citenamefont {Bednorz}\ and\ \citenamefont
  {M\"uller}(1986)}]{Bednorz}%
  \BibitemOpen
  \bibfield  {author} {\bibinfo {author} {\bibfnamefont {J.~G.}\ \bibnamefont
  {Bednorz}}\ and\ \bibinfo {author} {\bibfnamefont {K.~A.}\ \bibnamefont
  {M\"uller}},\ }\href@noop {} {\bibfield  {journal} {\bibinfo  {journal} {Z.
  Phys. B.}\ }\textbf {\bibinfo {volume} {64}},\ \bibinfo {pages} {189}
  (\bibinfo {year} {1986})}\BibitemShut {NoStop}%
\bibitem [{\citenamefont {Timusk}\ and\ \citenamefont {Statt}(1999)}]{Timusk}%
  \BibitemOpen
  \bibfield  {author} {\bibinfo {author} {\bibfnamefont {T.}~\bibnamefont
  {Timusk}}\ and\ \bibinfo {author} {\bibfnamefont {B.}~\bibnamefont {Statt}},\
  }\href@noop {} {\bibfield  {journal} {\bibinfo  {journal} {Rep. Prog. Phys.}\
  }\textbf {\bibinfo {volume} {62}},\ \bibinfo {pages} {61} (\bibinfo {year}
  {1999})}\BibitemShut {NoStop}%
\bibitem [{\citenamefont {Gao}\ \emph {et~al.}(1994)\citenamefont {Gao},
  \citenamefont {Xue}, \citenamefont {Chen}, \citenamefont {Xiong},
  \citenamefont {Meng}, \citenamefont {Ramirez}, \citenamefont {Chu},
  \citenamefont {Eggert},\ and\ \citenamefont {Mao}}]{Chu1994}%
  \BibitemOpen
  \bibfield  {author} {\bibinfo {author} {\bibfnamefont {L.}~\bibnamefont
  {Gao}}, \bibinfo {author} {\bibfnamefont {Y.~Y.}\ \bibnamefont {Xue}},
  \bibinfo {author} {\bibfnamefont {F.}~\bibnamefont {Chen}}, \bibinfo {author}
  {\bibfnamefont {Q.}~\bibnamefont {Xiong}}, \bibinfo {author} {\bibfnamefont
  {R.~L.}\ \bibnamefont {Meng}}, \bibinfo {author} {\bibfnamefont
  {D.}~\bibnamefont {Ramirez}}, \bibinfo {author} {\bibfnamefont {C.~W.}\
  \bibnamefont {Chu}}, \bibinfo {author} {\bibfnamefont {J.~H.}\ \bibnamefont
  {Eggert}}, \ and\ \bibinfo {author} {\bibfnamefont {H.~K.}\ \bibnamefont
  {Mao}},\ }\href@noop {} {\bibfield  {journal} {\bibinfo  {journal} {Phys.
  Rev. B}\ }\textbf {\bibinfo {volume} {50}},\ \bibinfo {pages} {4260(R)}
  (\bibinfo {year} {1994})}\BibitemShut {NoStop}%
\bibitem [{\citenamefont {Kamihara}\ \emph {et~al.}(2008)\citenamefont
  {Kamihara}, \citenamefont {Watanabe}, \citenamefont {Hirano},\ and\
  \citenamefont {Hosono}}]{Hosono}%
  \BibitemOpen
  \bibfield  {author} {\bibinfo {author} {\bibfnamefont {Y.}~\bibnamefont
  {Kamihara}}, \bibinfo {author} {\bibfnamefont {T.}~\bibnamefont {Watanabe}},
  \bibinfo {author} {\bibfnamefont {M.}~\bibnamefont {Hirano}}, \ and\ \bibinfo
  {author} {\bibfnamefont {H.}~\bibnamefont {Hosono}},\ }\href@noop {}
  {\bibfield  {journal} {\bibinfo  {journal} {J. Am. Chem. Soc.}\ }\textbf
  {\bibinfo {volume} {130}},\ \bibinfo {pages} {3296} (\bibinfo {year}
  {2008})}\BibitemShut {NoStop}%
\bibitem [{\citenamefont {Steglich}\ \emph {et~al.}(1979)\citenamefont
  {Steglich}, \citenamefont {Aarts}, \citenamefont {Bredl}, \citenamefont
  {Lieke}, \citenamefont {Meschede}, \citenamefont {Franz},\ and\ \citenamefont
  {Sch\"afer}}]{Steglich1979}%
  \BibitemOpen
  \bibfield  {author} {\bibinfo {author} {\bibfnamefont {F.}~\bibnamefont
  {Steglich}}, \bibinfo {author} {\bibfnamefont {J.}~\bibnamefont {Aarts}},
  \bibinfo {author} {\bibfnamefont {C.~D.}\ \bibnamefont {Bredl}}, \bibinfo
  {author} {\bibfnamefont {W.}~\bibnamefont {Lieke}}, \bibinfo {author}
  {\bibfnamefont {D.}~\bibnamefont {Meschede}}, \bibinfo {author}
  {\bibfnamefont {W.}~\bibnamefont {Franz}}, \ and\ \bibinfo {author}
  {\bibfnamefont {H.}~\bibnamefont {Sch\"afer}},\ }\href@noop {} {\bibfield
  {journal} {\bibinfo  {journal} {\prl}\ }\textbf {\bibinfo {volume} {43}},\
  \bibinfo {pages} {1892} (\bibinfo {year} {1979})}\BibitemShut {NoStop}%
\bibitem [{\citenamefont {McKenzie}(1997)}]{McKenzie}%
  \BibitemOpen
  \bibfield  {author} {\bibinfo {author} {\bibfnamefont {R.~H.}\ \bibnamefont
  {McKenzie}},\ }\href@noop {} {\bibfield  {journal} {\bibinfo  {journal}
  {Science}\ }\textbf {\bibinfo {volume} {278}},\ \bibinfo {pages} {820}
  (\bibinfo {year} {1997})}\BibitemShut {NoStop}%
\bibitem [{\citenamefont {Drozdov}\ \emph {et~al.}(2015)\citenamefont
  {Drozdov}, \citenamefont {Eremets}, \citenamefont {Troyan}, \citenamefont
  {Ksenofontov},\ and\ \citenamefont {Shylin}}]{H2S}%
  \BibitemOpen
  \bibfield  {author} {\bibinfo {author} {\bibfnamefont {A.}~\bibnamefont
  {Drozdov}}, \bibinfo {author} {\bibfnamefont {M.~I.}\ \bibnamefont
  {Eremets}}, \bibinfo {author} {\bibfnamefont {I.~A.}\ \bibnamefont {Troyan}},
  \bibinfo {author} {\bibfnamefont {V.}~\bibnamefont {Ksenofontov}}, \ and\
  \bibinfo {author} {\bibfnamefont {S.~I.}\ \bibnamefont {Shylin}},\
  }\href@noop {} {\bibfield  {journal} {\bibinfo  {journal} {Nature (London)}\
  }\textbf {\bibinfo {volume} {525}},\ \bibinfo {pages} {73} (\bibinfo {year}
  {2015})}\BibitemShut {NoStop}%
\bibitem [{\citenamefont {Wang}\ \emph {et~al.}(2012)\citenamefont {Wang},
  \citenamefont {Li}, \citenamefont {Zhang}, \citenamefont {Zhang},
  \citenamefont {Zhang}, \citenamefont {Li}, \citenamefont {Ding},
  \citenamefont {Ou}, \citenamefont {Deng}, \citenamefont {Chang},
  \citenamefont {Wen}, \citenamefont {Song}, \citenamefont {He}, \citenamefont
  {Jia}, \citenamefont {Ji}, \citenamefont {Wang}, \citenamefont {Wang},
  \citenamefont {Chen}, \citenamefont {Ma},\ and\ \citenamefont
  {Xue}}]{Xue2012}%
  \BibitemOpen
  \bibfield  {author} {\bibinfo {author} {\bibfnamefont {Q.-Y.}\ \bibnamefont
  {Wang}}, \bibinfo {author} {\bibfnamefont {Z.}~\bibnamefont {Li}}, \bibinfo
  {author} {\bibfnamefont {W.-H.}\ \bibnamefont {Zhang}}, \bibinfo {author}
  {\bibfnamefont {Z.-C.}\ \bibnamefont {Zhang}}, \bibinfo {author}
  {\bibfnamefont {J.-S.}\ \bibnamefont {Zhang}}, \bibinfo {author}
  {\bibfnamefont {W.}~\bibnamefont {Li}}, \bibinfo {author} {\bibfnamefont
  {H.}~\bibnamefont {Ding}}, \bibinfo {author} {\bibfnamefont {Y.-B.}\
  \bibnamefont {Ou}}, \bibinfo {author} {\bibfnamefont {P.}~\bibnamefont
  {Deng}}, \bibinfo {author} {\bibfnamefont {K.}~\bibnamefont {Chang}},
  \bibinfo {author} {\bibfnamefont {J.}~\bibnamefont {Wen}}, \bibinfo {author}
  {\bibfnamefont {C.-L.}\ \bibnamefont {Song}}, \bibinfo {author}
  {\bibfnamefont {K.}~\bibnamefont {He}}, \bibinfo {author} {\bibfnamefont
  {J.-F.}\ \bibnamefont {Jia}}, \bibinfo {author} {\bibfnamefont {S.-H.}\
  \bibnamefont {Ji}}, \bibinfo {author} {\bibfnamefont {Y.-Y.}\ \bibnamefont
  {Wang}}, \bibinfo {author} {\bibfnamefont {L.-L.}\ \bibnamefont {Wang}},
  \bibinfo {author} {\bibfnamefont {X.}~\bibnamefont {Chen}}, \bibinfo {author}
  {\bibfnamefont {X.-C.}\ \bibnamefont {Ma}}, \ and\ \bibinfo {author}
  {\bibfnamefont {Q.-K.}\ \bibnamefont {Xue}},\ }\href
  {http://stacks.iop.org/0256-307X/29/i=3/a=037402} {\bibfield  {journal}
  {\bibinfo  {journal} {Chin. Phys. Lett.}\ }\textbf {\bibinfo {volume} {29}},\
  \bibinfo {pages} {037402} (\bibinfo {year} {2012})}\BibitemShut {NoStop}%
\bibitem [{\citenamefont {Ge}\ \emph {et~al.}(2015)\citenamefont {Ge},
  \citenamefont {Liu}, \citenamefont {Liu}, \citenamefont {Gao}, \citenamefont
  {Qian}, \citenamefont {Xue}, \citenamefont {Liu},\ and\ \citenamefont
  {Jia}}]{Monolayer100K}%
  \BibitemOpen
  \bibfield  {author} {\bibinfo {author} {\bibfnamefont {J.-F.}\ \bibnamefont
  {Ge}}, \bibinfo {author} {\bibfnamefont {Z.-L.}\ \bibnamefont {Liu}},
  \bibinfo {author} {\bibfnamefont {C.}~\bibnamefont {Liu}}, \bibinfo {author}
  {\bibfnamefont {C.-L.}\ \bibnamefont {Gao}}, \bibinfo {author} {\bibfnamefont
  {D.}~\bibnamefont {Qian}}, \bibinfo {author} {\bibfnamefont {Q.-K.}\
  \bibnamefont {Xue}}, \bibinfo {author} {\bibfnamefont {Y.}~\bibnamefont
  {Liu}}, \ and\ \bibinfo {author} {\bibfnamefont {J.-F.}\ \bibnamefont
  {Jia}},\ }\href@noop {} {\bibfield  {journal} {\bibinfo  {journal} {Nat.
  Mater.}\ }\textbf {\bibinfo {volume} {14}},\ \bibinfo {pages} {285–289}
  (\bibinfo {year} {2015})}\BibitemShut {NoStop}%
\bibitem [{\citenamefont {Lee}(2015)}]{LeeDH}%
  \BibitemOpen
  \bibfield  {author} {\bibinfo {author} {\bibfnamefont {D.-H.}\ \bibnamefont
  {Lee}},\ }\href {http://stacks.iop.org/1674-1056/24/i=11/a=117405} {\bibfield
   {journal} {\bibinfo  {journal} {Chin. Phys. B}\ }\textbf {\bibinfo {volume}
  {24}},\ \bibinfo {pages} {117405} (\bibinfo {year} {2015})}\BibitemShut
  {NoStop}%
\bibitem [{\citenamefont {Cao}\ \emph {et~al.}(2018)\citenamefont {Cao},
  \citenamefont {Fatemi}, \citenamefont {Fang}, \citenamefont {Watanabe},
  \citenamefont {Taniguchi}, \citenamefont {Kaxiras},\ and\ \citenamefont
  {Jarillo-Herrero}}]{YCao1}%
  \BibitemOpen
  \bibfield  {author} {\bibinfo {author} {\bibfnamefont {Y.}~\bibnamefont
  {Cao}}, \bibinfo {author} {\bibfnamefont {V.}~\bibnamefont {Fatemi}},
  \bibinfo {author} {\bibfnamefont {S.}~\bibnamefont {Fang}}, \bibinfo {author}
  {\bibfnamefont {K.}~\bibnamefont {Watanabe}}, \bibinfo {author}
  {\bibfnamefont {T.}~\bibnamefont {Taniguchi}}, \bibinfo {author}
  {\bibfnamefont {E.}~\bibnamefont {Kaxiras}}, \ and\ \bibinfo {author}
  {\bibfnamefont {P.}~\bibnamefont {Jarillo-Herrero}},\ }\href@noop {}
  {\bibfield  {journal} {\bibinfo  {journal} {Nature (London)}\ }\textbf
  {\bibinfo {volume} {556}},\ \bibinfo {pages} {43–50} (\bibinfo {year}
  {2018})}\BibitemShut {NoStop}%
\bibitem [{\citenamefont {Uemura}(1997)}]{Uemura}%
  \BibitemOpen
  \bibfield  {author} {\bibinfo {author} {\bibfnamefont {Y.~J.}\ \bibnamefont
  {Uemura}},\ }\href@noop {} {\bibfield  {journal} {\bibinfo  {journal}
  {Physica C}\ }\textbf {\bibinfo {volume} {282-287}},\ \bibinfo {pages} {194}
  (\bibinfo {year} {1997})}\BibitemShut {NoStop}%
\bibitem [{\citenamefont {Chen}\ \emph {et~al.}(1999)\citenamefont {Chen},
  \citenamefont {Kosztin}, \citenamefont {Jank\'o},\ and\ \citenamefont
  {Levin}}]{Chen1}%
  \BibitemOpen
  \bibfield  {author} {\bibinfo {author} {\bibfnamefont {Q.~J.}\ \bibnamefont
  {Chen}}, \bibinfo {author} {\bibfnamefont {I.}~\bibnamefont {Kosztin}},
  \bibinfo {author} {\bibfnamefont {B.}~\bibnamefont {Jank\'o}}, \ and\
  \bibinfo {author} {\bibfnamefont {K.}~\bibnamefont {Levin}},\ }\href@noop {}
  {\bibfield  {journal} {\bibinfo  {journal} {Phys. Rev. B}\ }\textbf {\bibinfo
  {volume} {59}},\ \bibinfo {pages} {7083} (\bibinfo {year}
  {1999})}\BibitemShut {NoStop}%
\bibitem [{\citenamefont {Chen}\ \emph {et~al.}(2005)\citenamefont {Chen},
  \citenamefont {Stajic}, \citenamefont {Tan},\ and\ \citenamefont
  {Levin}}]{Review}%
  \BibitemOpen
  \bibfield  {author} {\bibinfo {author} {\bibfnamefont {Q.~J.}\ \bibnamefont
  {Chen}}, \bibinfo {author} {\bibfnamefont {J.}~\bibnamefont {Stajic}},
  \bibinfo {author} {\bibfnamefont {S.~N.}\ \bibnamefont {Tan}}, \ and\
  \bibinfo {author} {\bibfnamefont {K.}~\bibnamefont {Levin}},\ }\href@noop {}
  {\bibfield  {journal} {\bibinfo  {journal} {Phys. Rep.}\ }\textbf {\bibinfo
  {volume} {412}},\ \bibinfo {pages} {1} (\bibinfo {year} {2005})}\BibitemShut
  {NoStop}%
\bibitem [{\citenamefont {Nozi\`{e}res}\ and\ \citenamefont
  {Schmitt-Rink}(1985)}]{NSR}%
  \BibitemOpen
  \bibfield  {author} {\bibinfo {author} {\bibfnamefont {P.}~\bibnamefont
  {Nozi\`{e}res}}\ and\ \bibinfo {author} {\bibfnamefont {S.}~\bibnamefont
  {Schmitt-Rink}},\ }\href@noop {} {\bibfield  {journal} {\bibinfo  {journal}
  {J. Low Temp. Phys.}\ }\textbf {\bibinfo {volume} {59}},\ \bibinfo {pages}
  {195} (\bibinfo {year} {1985})}\BibitemShut {NoStop}%
\bibitem [{\citenamefont {Haussmann}\ \emph {et~al.}(2007)\citenamefont
  {Haussmann}, \citenamefont {Rantner}, \citenamefont {Cerrito},\ and\
  \citenamefont {Zwerger}}]{Zwerger}%
  \BibitemOpen
  \bibfield  {author} {\bibinfo {author} {\bibfnamefont {R.}~\bibnamefont
  {Haussmann}}, \bibinfo {author} {\bibfnamefont {W.}~\bibnamefont {Rantner}},
  \bibinfo {author} {\bibfnamefont {S.}~\bibnamefont {Cerrito}}, \ and\
  \bibinfo {author} {\bibfnamefont {W.}~\bibnamefont {Zwerger}},\ }\href@noop
  {} {\bibfield  {journal} {\bibinfo  {journal} {\pra}\ }\textbf {\bibinfo
  {volume} {75}},\ \bibinfo {pages} {023610} (\bibinfo {year}
  {2007})}\BibitemShut {NoStop}%
\bibitem [{\citenamefont {Burovski}\ \emph {et~al.}(2008)\citenamefont
  {Burovski}, \citenamefont {Kozik}, \citenamefont {Prokof'ev}, \citenamefont
  {Svistunov},\ and\ \citenamefont {Troyer}}]{TroyerPRL2008}%
  \BibitemOpen
  \bibfield  {author} {\bibinfo {author} {\bibfnamefont {E.}~\bibnamefont
  {Burovski}}, \bibinfo {author} {\bibfnamefont {E.}~\bibnamefont {Kozik}},
  \bibinfo {author} {\bibfnamefont {N.}~\bibnamefont {Prokof'ev}}, \bibinfo
  {author} {\bibfnamefont {B.}~\bibnamefont {Svistunov}}, \ and\ \bibinfo
  {author} {\bibfnamefont {M.}~\bibnamefont {Troyer}},\ }\href@noop {}
  {\bibfield  {journal} {\bibinfo  {journal} {\prl}\ }\textbf {\bibinfo
  {volume} {101}},\ \bibinfo {pages} {090402} (\bibinfo {year}
  {2008})}\BibitemShut {NoStop}%
\bibitem [{\citenamefont {Goulko}\ and\ \citenamefont
  {Wingate}(2010)}]{Wingate}%
  \BibitemOpen
  \bibfield  {author} {\bibinfo {author} {\bibfnamefont {O.}~\bibnamefont
  {Goulko}}\ and\ \bibinfo {author} {\bibfnamefont {M.}~\bibnamefont
  {Wingate}},\ }\href@noop {} {\bibfield  {journal} {\bibinfo  {journal} {Phys.
  Rev. A}\ }\textbf {\bibinfo {volume} {82}},\ \bibinfo {pages} {053621}
  (\bibinfo {year} {2010})}\BibitemShut {NoStop}%
\bibitem [{\citenamefont {Bulgac}\ \emph {et~al.}(2006)\citenamefont {Bulgac},
  \citenamefont {Drut},\ and\ \citenamefont {Magierski}}]{BulgacMC}%
  \BibitemOpen
  \bibfield  {author} {\bibinfo {author} {\bibfnamefont {A.}~\bibnamefont
  {Bulgac}}, \bibinfo {author} {\bibfnamefont {J.}~\bibnamefont {Drut}}, \ and\
  \bibinfo {author} {\bibfnamefont {P.}~\bibnamefont {Magierski}},\ }\href@noop
  {} {\bibfield  {journal} {\bibinfo  {journal} {Phys. Rev. Lett.}\ }\textbf
  {\bibinfo {volume} {96}},\ \bibinfo {pages} {090404} (\bibinfo {year}
  {2006})},\ \bibinfo {note} {\textit{ibid.} \textbf{99}, 120401
  (2007).}\BibitemShut {Stop}%
\bibitem [{\citenamefont {Akkineni}\ \emph {et~al.}(2007)\citenamefont
  {Akkineni}, \citenamefont {Ceperley},\ and\ \citenamefont
  {Trivedi}}]{Akkineni}%
  \BibitemOpen
  \bibfield  {author} {\bibinfo {author} {\bibfnamefont {V.~K.}\ \bibnamefont
  {Akkineni}}, \bibinfo {author} {\bibfnamefont {D.~M.}\ \bibnamefont
  {Ceperley}}, \ and\ \bibinfo {author} {\bibfnamefont {N.}~\bibnamefont
  {Trivedi}},\ }\href@noop {} {\bibfield  {journal} {\bibinfo  {journal} {Phys.
  Rev. B}\ }\textbf {\bibinfo {volume} {76}},\ \bibinfo {pages} {165116}
  (\bibinfo {year} {2007})}\BibitemShut {NoStop}%
\bibitem [{\citenamefont {Floerchinger}\ \emph {et~al.}(2008)\citenamefont
  {Floerchinger}, \citenamefont {Scherer}, \citenamefont {Diehl},\ and\
  \citenamefont {Wetterich}}]{Floerchinger}%
  \BibitemOpen
  \bibfield  {author} {\bibinfo {author} {\bibfnamefont {S.}~\bibnamefont
  {Floerchinger}}, \bibinfo {author} {\bibfnamefont {M.}~\bibnamefont
  {Scherer}}, \bibinfo {author} {\bibfnamefont {S.}~\bibnamefont {Diehl}}, \
  and\ \bibinfo {author} {\bibfnamefont {C.}~\bibnamefont {Wetterich}},\
  }\href@noop {} {\bibfield  {journal} {\bibinfo  {journal} {\prb}\ }\textbf
  {\bibinfo {volume} {78}},\ \bibinfo {pages} {174528} (\bibinfo {year}
  {2008})}\BibitemShut {NoStop}%
\bibitem [{\citenamefont {Kinast}\ \emph {et~al.}(2005)\citenamefont {Kinast},
  \citenamefont {Turlapov}, \citenamefont {Thomas}, \citenamefont {Chen},
  \citenamefont {Stajic},\ and\ \citenamefont {Levin}}]{ThermoScience}%
  \BibitemOpen
  \bibfield  {author} {\bibinfo {author} {\bibfnamefont {J.}~\bibnamefont
  {Kinast}}, \bibinfo {author} {\bibfnamefont {A.}~\bibnamefont {Turlapov}},
  \bibinfo {author} {\bibfnamefont {J.~E.}\ \bibnamefont {Thomas}}, \bibinfo
  {author} {\bibfnamefont {Q.~J.}\ \bibnamefont {Chen}}, \bibinfo {author}
  {\bibfnamefont {J.}~\bibnamefont {Stajic}}, \ and\ \bibinfo {author}
  {\bibfnamefont {K.}~\bibnamefont {Levin}},\ }\href@noop {} {\bibfield
  {journal} {\bibinfo  {journal} {Science}\ }\textbf {\bibinfo {volume}
  {307}},\ \bibinfo {pages} {1296} (\bibinfo {year} {2005})}\BibitemShut
  {NoStop}%
\bibitem [{\citenamefont {Ku}\ \emph {et~al.}(2012)\citenamefont {Ku},
  \citenamefont {Sommer}, \citenamefont {Cheuk},\ and\ \citenamefont
  {Zwierlein}}]{Zwierlein2012}%
  \BibitemOpen
  \bibfield  {author} {\bibinfo {author} {\bibfnamefont {M.~J.~H.}\
  \bibnamefont {Ku}}, \bibinfo {author} {\bibfnamefont {A.~T.}\ \bibnamefont
  {Sommer}}, \bibinfo {author} {\bibfnamefont {L.~W.}\ \bibnamefont {Cheuk}}, \
  and\ \bibinfo {author} {\bibfnamefont {M.~W.}\ \bibnamefont {Zwierlein}},\
  }\href@noop {} {\bibfield  {journal} {\bibinfo  {journal} {Science}\ }\textbf
  {\bibinfo {volume} {335}},\ \bibinfo {pages} {563} (\bibinfo {year}
  {2012})}\BibitemShut {NoStop}%
\bibitem [{\citenamefont {Lamporesi}\ \emph {et~al.}(2010)\citenamefont
  {Lamporesi}, \citenamefont {Catani}, \citenamefont {Barontini}, \citenamefont
  {Nishida}, \citenamefont {Inguscio},\ and\ \citenamefont
  {Minardi}}]{Lamporesi10PRL}%
  \BibitemOpen
  \bibfield  {author} {\bibinfo {author} {\bibfnamefont {G.}~\bibnamefont
  {Lamporesi}}, \bibinfo {author} {\bibfnamefont {J.}~\bibnamefont {Catani}},
  \bibinfo {author} {\bibfnamefont {G.}~\bibnamefont {Barontini}}, \bibinfo
  {author} {\bibfnamefont {Y.}~\bibnamefont {Nishida}}, \bibinfo {author}
  {\bibfnamefont {M.}~\bibnamefont {Inguscio}}, \ and\ \bibinfo {author}
  {\bibfnamefont {F.}~\bibnamefont {Minardi}},\ }\href {\doibase
  10.1103/PhysRevLett.104.153202} {\bibfield  {journal} {\bibinfo  {journal}
  {Phys. Rev. Lett.}\ }\textbf {\bibinfo {volume} {104}},\ \bibinfo {pages}
  {153202} (\bibinfo {year} {2010})}\BibitemShut {NoStop}%
\bibitem [{\citenamefont {Iskin}\ and\ \citenamefont {Suba\ifmmode
  \mbox{\c{s}}\else \c{s}\fi{}\ifmmode \imath \else~\i
  \fi{}}(2010)}]{Iskin10PRA}%
  \BibitemOpen
  \bibfield  {author} {\bibinfo {author} {\bibfnamefont {M.}~\bibnamefont
  {Iskin}}\ and\ \bibinfo {author} {\bibfnamefont {A.~L.}\ \bibnamefont
  {Suba\ifmmode \mbox{\c{s}}\else \c{s}\fi{}\ifmmode \imath \else~\i \fi{}}},\
  }\href {\doibase 10.1103/PhysRevA.82.063628} {\bibfield  {journal} {\bibinfo
  {journal} {Phys. Rev. A}\ }\textbf {\bibinfo {volume} {82}},\ \bibinfo
  {pages} {063628} (\bibinfo {year} {2010})}\BibitemShut {NoStop}%
\bibitem [{\citenamefont {Yang}\ \emph {et~al.}(2011)\citenamefont {Yang},
  \citenamefont {Huang},\ and\ \citenamefont {Wan}}]{XYang11EPJB}%
  \BibitemOpen
  \bibfield  {author} {\bibinfo {author} {\bibfnamefont {X.~S.}\ \bibnamefont
  {Yang}}, \bibinfo {author} {\bibfnamefont {B.~B.}\ \bibnamefont {Huang}}, \
  and\ \bibinfo {author} {\bibfnamefont {S.~L.}\ \bibnamefont {Wan}},\ }\href
  {\doibase 10.1140/epjb/e2011-20354-0} {\bibfield  {journal} {\bibinfo
  {journal} {Eur. Phys. J. B}\ }\textbf {\bibinfo {volume} {83}},\ \bibinfo
  {pages} {445} (\bibinfo {year} {2011})}\BibitemShut {NoStop}%
\bibitem [{\citenamefont {Zhang}\ \emph {et~al.}(2017)\citenamefont {Zhang},
  \citenamefont {Che}, \citenamefont {Wang},\ and\ \citenamefont
  {Chen}}]{MixedDim_Srep}%
  \BibitemOpen
  \bibfield  {author} {\bibinfo {author} {\bibfnamefont {L.~F.}\ \bibnamefont
  {Zhang}}, \bibinfo {author} {\bibfnamefont {Y.~M.}\ \bibnamefont {Che}},
  \bibinfo {author} {\bibfnamefont {J.~B.}\ \bibnamefont {Wang}}, \ and\
  \bibinfo {author} {\bibfnamefont {Q.~J.}\ \bibnamefont {Chen}},\ }\href@noop
  {} {\bibfield  {journal} {\bibinfo  {journal} {Sci. Rep.}\ }\textbf {\bibinfo
  {volume} {7}},\ \bibinfo {pages} {12948} (\bibinfo {year}
  {2017})}\BibitemShut {NoStop}%
\bibitem [{\citenamefont {Fulde}\ and\ \citenamefont {Ferrell}(1964)}]{FF}%
  \BibitemOpen
  \bibfield  {author} {\bibinfo {author} {\bibfnamefont {P.}~\bibnamefont
  {Fulde}}\ and\ \bibinfo {author} {\bibfnamefont {R.~A.}\ \bibnamefont
  {Ferrell}},\ }\href {\doibase 10.1103/PhysRev.135.A550} {\bibfield  {journal}
  {\bibinfo  {journal} {Phys. Rev.}\ }\textbf {\bibinfo {volume} {135}},\
  \bibinfo {pages} {A550} (\bibinfo {year} {1964})}\BibitemShut {NoStop}%
\bibitem [{\citenamefont {Larkin}\ and\ \citenamefont
  {Ovchinnikov}(1965)}]{LO}%
  \BibitemOpen
  \bibfield  {author} {\bibinfo {author} {\bibfnamefont {A.~I.}\ \bibnamefont
  {Larkin}}\ and\ \bibinfo {author} {\bibfnamefont {Y.~N.}\ \bibnamefont
  {Ovchinnikov}},\ }\href@noop {} {\bibfield  {journal} {\bibinfo  {journal}
  {Sov. Phys. JETP}\ }\textbf {\bibinfo {volume} {20}},\ \bibinfo {pages} {762}
  (\bibinfo {year} {1965})},\ \bibinfo {note} {[Zh. Eksp. Teor. Fiz.
  \textbf{47}, 1136 (1964)]}\BibitemShut {NoStop}%
\bibitem [{\citenamefont {Wang}\ \emph {et~al.}(2018)\citenamefont {Wang},
  \citenamefont {Che}, \citenamefont {Zhang},\ and\ \citenamefont
  {Chen}}]{FFLO_Instability}%
  \BibitemOpen
  \bibfield  {author} {\bibinfo {author} {\bibfnamefont {J.~B.}\ \bibnamefont
  {Wang}}, \bibinfo {author} {\bibfnamefont {Y.~M.}\ \bibnamefont {Che}},
  \bibinfo {author} {\bibfnamefont {L.~F.}\ \bibnamefont {Zhang}}, \ and\
  \bibinfo {author} {\bibfnamefont {Q.~J.}\ \bibnamefont {Chen}},\ }\href@noop
  {} {\bibfield  {journal} {\bibinfo  {journal} {\prb}\ }\textbf {\bibinfo
  {volume} {97}},\ \bibinfo {pages} {134513} (\bibinfo {year}
  {2018})}\BibitemShut {NoStop}%
\bibitem [{\citenamefont {Chen}\ \emph {et~al.}(1998)\citenamefont {Chen},
  \citenamefont {Kosztin}, \citenamefont {Jank\'o},\ and\ \citenamefont
  {Levin}}]{Chen2}%
  \BibitemOpen
  \bibfield  {author} {\bibinfo {author} {\bibfnamefont {Q.~J.}\ \bibnamefont
  {Chen}}, \bibinfo {author} {\bibfnamefont {I.}~\bibnamefont {Kosztin}},
  \bibinfo {author} {\bibfnamefont {B.}~\bibnamefont {Jank\'o}}, \ and\
  \bibinfo {author} {\bibfnamefont {K.}~\bibnamefont {Levin}},\ }\href@noop {}
  {\bibfield  {journal} {\bibinfo  {journal} {Phys. Rev. Lett.}\ }\textbf
  {\bibinfo {volume} {81}},\ \bibinfo {pages} {4708} (\bibinfo {year}
  {1998})}\BibitemShut {NoStop}%
\bibitem [{not({\natexlab{a}})}]{noteonm_eff}%
  \BibitemOpen
  \href@noop {} {} \bibinfo {note} {Here the lattice
  component of the two pairing atoms contributes $1/m_z\sim 2td^2$, the other 5
  components each contribute $1/m$, and $1/m_{eff}$ is given by the
  average.}\BibitemShut {Stop}%
\bibitem [{not({\natexlab{a}})}]{noteonlarge_td2}%
  \BibitemOpen
  \href@noop {} {} \bibinfo {note} {The $2mtd^2 > 1 $ regime
  may become accessible via introducing an attractive interaction within the
  lattice component, especially in the lattice direction, via, e.g, electric
  dipole-dipole interactions and so on.}\BibitemShut {Stop}%
\bibitem [{not({\natexlab{b}})}]{noteonmassimb}%
  \BibitemOpen
  \href@noop {} {} \bibinfo {note} {This can be understood
  from the existence of a finite $T_c$ as $1/k_Fa$ goes all the way down to
  $-\infty$ in the pure 3D continuum \protect\cite{Guo2009PRA}.}\BibitemShut
  {Stop}%
\bibitem [{\citenamefont {Bickers}\ \emph {et~al.}(1989)\citenamefont
  {Bickers}, \citenamefont {Scalapino},\ and\ \citenamefont {White}}]{Bickers}%
  \BibitemOpen
  \bibfield  {author} {\bibinfo {author} {\bibfnamefont {N.~E.}\ \bibnamefont
  {Bickers}}, \bibinfo {author} {\bibfnamefont {D.~J.}\ \bibnamefont
  {Scalapino}}, \ and\ \bibinfo {author} {\bibfnamefont {S.~R.}\ \bibnamefont
  {White}},\ }\href@noop {} {\bibfield  {journal} {\bibinfo  {journal} {Phys.
  Rev. Lett.}\ }\textbf {\bibinfo {volume} {62}},\ \bibinfo {pages} {961}
  (\bibinfo {year} {1989})}\BibitemShut {NoStop}%
\bibitem [{\citenamefont {Guo}\ \emph {et~al.}(2009)\citenamefont {Guo},
  \citenamefont {Chien}, \citenamefont {Chen}, \citenamefont {He},\ and\
  \citenamefont {Levin}}]{Guo2009PRA}%
  \BibitemOpen
  \bibfield  {author} {\bibinfo {author} {\bibfnamefont {H.}~\bibnamefont
  {Guo}}, \bibinfo {author} {\bibfnamefont {C.-C.}\ \bibnamefont {Chien}},
  \bibinfo {author} {\bibfnamefont {Q.~J.}\ \bibnamefont {Chen}}, \bibinfo
  {author} {\bibfnamefont {Y.}~\bibnamefont {He}}, \ and\ \bibinfo {author}
  {\bibfnamefont {K.}~\bibnamefont {Levin}},\ }\href {\doibase
  10.1103/PhysRevA.80.011601} {\bibfield  {journal} {\bibinfo  {journal} {Phys.
  Rev. A}\ }\textbf {\bibinfo {volume} {80}},\ \bibinfo {pages} {011601(R)}
  (\bibinfo {year} {2009})}\BibitemShut {NoStop}%
\end{thebibliography}
%

\end{document}